\documentclass[aps,prl,twocolumn,showpacs]{revtex4}
\usepackage{graphicx}
\usepackage{epsf}
\begin{document}
\title{Bogoliubov shadow bands in the normal state \\of superconducting 
systems 
 with strong pair fluctuations}

\author{T. ~Domanski$^{(a,b)}$ and J.~Ranninger$^{(a)}$}
\affiliation{${(a)}$ 
Centre de Recherches sur les Tr\`es Basses Temp\'eratures
 associ\'e \`a l'Universit\'e Joseph Fourier,
 C.N.R.S., BP 166, 38042 Grenoble-C\'edex 9, France}

\affiliation{${(b)}$ 
 Institute of Physics, M.\ Curie Sk\l odowska University, 
 20-031 Lublin, Poland}

\begin{abstract}

On the basis of a scenario where electron pairing is induced by resonant 
two-particle scattering (the Boson Fermion model), we show how precursors 
of the superconducting state - in form of overdamped Bogoliubov modes - emerge 
in the normal state upon approaching the transition temperature from above. 
This result is obtained by a renormalization technique based on continuous 
unitary transformations (the flow equations), projecting out the coherent 
contributions in the electron spectral function from an incoherent background.

\end{abstract}
\maketitle

The discovery of high temperature superconductivity (HTS) and the enormous 
theoretical effort following this discovery has led to reconsider the whole 
issue of superconductivity and to put it into a more general and 
broader perspective than that of the familiar BCS scenario. 
It became clear early on that in these new 
superconductors one is confronted with a 
situation between that of classical Cooper pairing (controlled by the 
amplitude of the order parameter) and that of a superfluid phase of 
tightly bound electron pairs (controlled by the phase fluctuation). 
The theoretical issue is to understand the interplay between the two, 
which amounts to fully taking into account the interaction between 
single electron states 
and  electron pair fluctuations. A particularly clear presentation of this 
problematic was given  by Tchernyshyov \cite{Tchernyshyov-97}, who introduced 
an effective amplitude for electron pairs in the normal state by 
including in the propagator of an electron a process involving the 
simultaneous propagation of a hole together with a Cooperon. This leads to the 
emergence of two-electron resonant states (long lived pair fluctuations) 
inside the Fermi sea, manifest in form of a pseudogap - 
a precursor of the superconducting gap in the normal state. 
The phenomenon of superconducting correlations building up above $T_c$ - 
the onset temperature of superconductivity - is by no means restricted to 
the HTS. It is encountered in a great variety of physical systems, where 
resonance pairing of fermionic quasi particle plays a key role and which can 
be discussed on the basis of the phenomenological Boson Fermion 
model (BFM). This model describes the formation of resonance bound 
Fermion-pairs inside the Fermi sea of uncorrelated Fermions.
The resonant bound states are  monitored by an exchange interaction with 
localized Bosons (tightly bound Fermion pairs, the specific origin of which 
will be of no importance for the present study). 
Realizations of such physics are be found in:
i) electron-phonon coupled systems in the intermediary coupling 
limit (see for instance ref.\cite{Ranninger-02} for which this BFM
was devised initially) describing an exchange interaction between localized 
bipolarons and itinerant electrons,
ii) the positive U Hubbard model\cite{Auerbach-02} describing an exchange 
interaction between spinon singlets of RVB electron pairs and holons,
iii) low density nuclear matter with isospin singlet pairing\cite{Schnell-99} 
and iv) the Feshbach resonance in atomic 
physics\cite{Timmermans-99,Domanski-03} 
invoked for tuning a quasi-bound state of atoms through a threshold resulting 
in a resonance superfluidity in traps\cite{Chiofalo-02,Ohashi-03}, and which 
involves entanglements of atoms in squeezed  states\cite{Yurovsky-03}. 
 
A salient feature of this BFM is the appearance of a pseudogap in the 
normal state single Fermion density of states when decreasing the 
temperature below a certain value $T^*$. Accompanied with this is a 
qualitative change in the transport: going from single Fermion transport 
above $T^*$ (describing a relatively good metal) to one involving well 
defined itinerant bosonic charge carriers (describing a bosonic metal 
coexisting with a poor fermionic metal\cite{Robin-98,Devillard-00}). Below 
a certain critical temperature $T_c$, the itinerant Bosons together with  
finite momentum Cooperons, induced in the fermionic subsystem, end up in a 
phase correlated superfluid state. One hence should not only expect a 
pseudogap in the Fermionic density of states as a precursor to the 
superconducting phase, but also corresponding modifications of the sizable 
incoherent component of the Fermionic spectral functions, which characterizes 
the pseudogap phase in the temperature interval $[T_c,T^*]$, to evolve into 
damped, but nevertheless clearly discernible,  Bogoliubov modes as one 
approaches $T_c$ . The feasibility of such 
a scenario will be the issue of the present study.

Applying a flow equation renormalization technique to the BFM  permits 
us to construct renormalized Fermionic operators which posses features 
of potentially containing residues of Bogoliubov type modes in the normal 
state. The essence of this technique consists in devising an infinite series 
of continuous unitary transformations\cite{Wegner-94,Glazek-94} for the 
Hamiltonian, such as  $H(l)=e^{S(l)}He^{-S(l)}$, where $l$ denotes the 
continuous flow parameter. Imposing a constraint structure on the 
renormalized Hamiltonian defines  renormalization 
equations for the parameters of this Hamiltonian which are 
devised in such a way that the exchange interaction between the Bosons and 
the Fermions is renormalized to zero giving rise to decoupled 
systems of Bosons and itinerant Fermions, whose parameters however 
have been rendered interdependent in the course of such a 
renormalization procedure. In the momentum representation this Hamiltonian 
consists of two parts $H(l)=H_{0}(l)+H_{int}(l)$ and has the following 
structure in the $l$-th step of this continuous transformation
\begin{eqnarray}
H_{0}(l)&=&\sum_{{\bf k},\sigma}(\varepsilon_{\bf k}(l)
-\mu) c_{{\bf k}\sigma}^{\dagger}c_{{\bf k}\sigma}+
\sum_{\bf q} (E_{\bf q}(l)-2\mu) b_{\bf q}^{\dagger}b_{\bf q} 
 \nonumber \\ &+&  \frac{1}{N}
\sum_{{\bf k},{\bf p},{\bf q}} U_{{\bf k},{\bf p},{\bf q}}(l)
c_{{\bf k}\uparrow}^{\dagger} c_{{\bf p}\downarrow}^{\dagger}
 c_{{\bf q}\downarrow} c_{{\bf k}+{\bf p}-{\bf q}\uparrow}
\label{H_0} \\   
H_{int}(l)&=&\frac{1}{\sqrt{N}}\sum_{{\bf k},{\bf p}} 
v_{{\bf k},{\bf p}}(l) \left(  b_{{\bf p}+{\bf k}}^{\dagger}
c_{{\bf k}\downarrow} c_{{\bf p}\uparrow} +
\mbox{h.c.} \right).
\label{H_pert}
\end{eqnarray}
$c_{{\bf k}\sigma}^{(\dagger)}$  refer to annihilation 
(creation) operators for the itinerant Fermions
with the energy $\varepsilon_{\bf k}(l)$ and 
$b_{\bf q}^{\dagger}$ ($b_{\bf q}$) denote boson operators representing 
bound Fermion pairs with energy $E_{\bf q}(l)-2\mu$. The 
Boson-Fermion exchange coupling is denoted by $v_{{\bf k},{\bf p}}(l)$ and 
the interaction  between Fermions by $U_{{\bf k},{\bf p},{\bf q}}(l)$. For 
$l=0$ these parameters reduce to the bare quantities 
$\varepsilon_{\bf k}-\mu$, $\Delta_B-2\mu$, $v$ and $0$ respectively, which 
characterize the initial Hamiltonian.   

In some recent work\cite{Domanski-01} we investigated the structure of 
the final renormalized Hamiltonian. Its resulting energy spectrum and 
single Fermion density of states showed the opening up of a pseudogap below 
a certain $T^*$ of the order of $v$. We here shall investigate the 
spectral function of these Fermions and how it changes as we go from the 
pseudogap phase $[T_c,T^*]$ into the superconducting one below $T_c$. 
We propose for that purpose a procedure for evaluating correlation functions 
within such a flow equation technique, which closely follows the standard 
procedure for renormalizing the Hamiltonian, determined by the  
differential equation:
\begin{eqnarray}
dH(l)/dl = [ \eta(l),H(l) ], 
\label{Hflow}
\end{eqnarray}  
and subject to the initial condition $H(0)$, presenting the original 
Hamiltonian. A suitably generating operator 
\begin{eqnarray}
\eta(l)= - \frac{1}{\sqrt{N}} \sum_{{\bf k},{\bf p}}
\alpha_{{\bf k},{\bf p}}(l) \left(  b_{{\bf p}+{\bf k}}^{\dagger}
c_{{\bf k}\downarrow} c_{{\bf p}\uparrow} - 
\mbox{h.c.} \right) 
\label{eta}
\end{eqnarray} 
where $\alpha_{{\bf k},{\bf p}}(l)= \left( \varepsilon_{\bf k}(l)
+\varepsilon_{\bf p}(l)-E_{{\bf k}+{\bf p}}(l) \right)  v_{{\bf k},
{\bf p}}(l)$ is chosen in such a way\cite{Wegner-94} that 
$\lim_{l\rightarrow\infty}H_{int}(l)=0$. This leads to a set of differential 
equations (given by eqs. (16-21) in Ref. \cite{Domanski-01}) which determines 
the evolution of the $l$ dependent parameters of the Hamiltonian. For 
consistency reasons, a flow equation procedure for an arbitrary operator 
$O(l)$ ought to be controlled by a formally equivalent equation of that 
defining the evolution of the Hamiltonian, i.e., eq. \ref{Hflow} 
where $H(l)$ is replaced by $O(l)$.  
This leads to a flow parameter dependent parameterization of the various 
operators for which we impose (similar to the procedure for deriving a 
renormalized Hamiltonian) the following constraint structure for the 
fermion operators:
\begin{eqnarray}
{c_{{\bf k}\uparrow}(l) \choose c_{-{\bf k}\downarrow}^{\dagger}(l)} =   
{{\cal P}_{\bf k}(l) \choose -{\cal R}_{\bf k}^{*}(l)} \;
c_{{\bf k}\uparrow} + {{\cal R}_{\bf k}(l) \choose {\cal P}_{\bf k}^{*}(l)} \; 
c_{-{\bf k}\downarrow}^{\dagger} \;\;\;\;\;\;\;\;  
 \nonumber \\
 +  \; \frac{1}{\sqrt{N}} 
\sum_{{\bf q} \neq{\bf 0}} \left[
{p_{{\bf k},{\bf q}}(l) \choose r_{{\bf k},{\bf q}}^{*}(l)} \; 
b_{\bf q}^{\dagger}
c_{{\bf q}+{\bf k}\uparrow}  + 
{r_{{\bf k},{\bf q}}(l) \choose -p_{{\bf k},{\bf q}}^{*}(l)} \; b_{\bf q} 
c_{{\bf q}-{\bf k}\downarrow}^{\dagger}
\right].
\label{Ansatz}
\end{eqnarray}
which generalizes the standard Bogoliubov transformation in two ways:
(a) the initial particle and hole operators are transformed
into Bogoliubov modes in a continuous way, 
(b) the correlated motion involving Fermion holes and Cooperons 
is taken into account via the terms proportional to 
$p_{{\bf k},{\bf q}}(l)$ and $r_{{\bf k},{\bf q}}(l)$.

These flow dependent parameters are then determined by the following set of 
differential equations: 
\begin{eqnarray}
 \frac{d{\cal P}_{\bf k}(l)}{dl} & = &  
\sqrt{n_{cond}^{B}} \; \alpha_{-{\bf k},{\bf k}}(l)
\; {\cal R}_{\bf k}(l) \label{P_flow} \\
& + & \frac{1}{N} \sum_{{\bf q}\neq{\bf 0}} 
\alpha_{{\bf q}-{\bf k},{\bf k}}(l) \left( 
n_{\bf q}^{B} + n_{{\bf q}-{\bf k}\downarrow}^{F}
\right) r_{{\bf k},{\bf q}}(l)
\nonumber \\ 
\frac{d{\cal R}_{\bf k}(l)}{dl} & = & - \;
\sqrt{n_{cond}^{B}} \; \alpha_{{\bf k},-{\bf k}}(l)
\; {\cal P}_{\bf k}(l) \label{R_flow} \\
& - & \frac{1}{N} \sum_{{\bf q}\neq{\bf 0}}
\alpha_{-{\bf k},{\bf q}+{\bf k}}(l) \left(
n_{\bf q}^{B} + n_{{\bf q}+{\bf k}\uparrow}^{F}
\right) p_{{\bf k},{\bf q}}(l)
\nonumber \\ 
\frac{d p_{{\bf k},{\bf q}}(l)}{dl} & = &
\alpha_{-{\bf k},{\bf q}+{\bf k}}(l) \; {\cal R}_{\bf k}(l)
 \label{p_flow} \\
 \frac{d r_{{\bf k},{\bf q}}(l)}{dl} & = & - \;
\alpha_{{\bf k},{\bf q}-{\bf k}}(l)  {\cal P}_{\bf k}(l)
\label{r_flow}
\end{eqnarray}
with the initial conditions ${\cal P}_{\bf k}
(0)=1$, ${\cal R}_{\bf k}(0)=0$, $p_{{\bf k},{\bf q}}(0)=0$,
$r_{{\bf k},{\bf q}}(0)=0$. $n_{cond}^{B}$ denotes the fraction of 
the condensed bosons and $n_{\bf q}^{B} 
\equiv \langle b_{\bf q}^{\dagger}b_{\bf q} \rangle 
=[\mbox{exp}(\tilde{E}_{\bf q}/k_{B}T)-1]^{-1}$ 
the distribution of the finite momentum 
(${\bf q}\neq{\bf 0}$) bosons. This set of equations has to be solved in 
conjunction with those (eqs 16-21 in ref. 14) determining the evolution 
of the various parameters entering the Hamiltonian ($\varepsilon_{\bf k}(l)$, 
$E_q(l)$ and $U_{\bf k,p,q}(l)$) being linked together via 
the expression for $\alpha_{\bf k,p}(l)$. The above flow equations 
(\ref{P_flow}-\ref{r_flow}) satisfy the sum rule 
\begin{eqnarray}
1 & = & | {\cal P}_{\bf k}(l)|^{2} + 
\frac{1}{N} \sum_{{\bf q}\neq{\bf 0}} \left( 
n_{\bf q}^{B} + n_{{\bf q}+{\bf k}\uparrow}^{F}
\right) | p_{{\bf k},{\bf q}}(l) |^{2} 
\nonumber \\ & + &
| {\cal R}_{\bf k}(l)|^{2} +
\frac{1}{N} \sum_{{\bf q}\neq{\bf 0}} \left(
n_{\bf q}^{B} + n_{{\bf q}-{\bf k}\downarrow}^{F}
\right) | r_{{\bf k},{\bf q}}(l) |^{2}, 
\label{PQ_inv}
\end{eqnarray}
relating the weights of the various coherent and incoherent contributions 
to the spectral function.
With these definitions, the single Fermion spectral function becomes
\begin{eqnarray}
A^{F}({\bf k},\omega)   
=| {\cal P}_{\bf k}(\infty) |^{2} \delta \left(
\omega - \tilde{\varepsilon}_{\bf k} \right)
+ | {\cal R}_{\bf k}(\infty)|^{2} \delta \left( 
\omega + \tilde{\varepsilon}_{-{\bf k}} \right)
\nonumber \\
+ \frac{1}{N} \sum_{{\bf q}\neq{\bf 0}} 
\left( n_{\bf q}^{B} + n_{{\bf q}+{\bf k}
\uparrow}^{F} \right) | p_{{\bf k},{\bf q}}
(\infty) |^{2} \delta ( \omega + \tilde{E}_{\bf q}
- \tilde{\varepsilon}_{{\bf q}+{\bf k}} ) \qquad 
\nonumber \\
+ \frac{1}{N} \sum_{{\bf q}\neq{\bf 0}}  
\left( n_{\bf q}^{B} + n_{{\bf q}-{\bf k}
\downarrow}^{F} \right) | r_{{\bf k},{\bf q}} 
(\infty) |^{2} \delta ( \omega - \tilde{E}_{\bf q}
+ \tilde{\varepsilon}_{{\bf q}-{\bf k}} ) \qquad 
\label{f_spectral} 
\end{eqnarray}
which is composed of a coherent part $A_{coh}^{F}({\bf k},\omega)$,
represented by a $\delta$-function like peak, and a remaining incoherent 
background 
$A_{inc}^{F}({\bf k})$, given by the last two terms in eq. (\ref{f_spectral}).
$\tilde{\varepsilon}_{\bf k}$ and $\tilde{E}_{\bf q}$ refer to the end 
results of the renormalization procedure ($l=\infty$) for the Fermion and, 
respectively, Boson spectra.  
We now analyze the structure of (\ref{f_spectral}) for different 
characteristic temperature regimes: $T>T^*$, the pseudogap regime 
$[T_c,T^*]$ and the superconducting regime $T<T_c$.  

\begin{figure}

\centerline{\epsfxsize=8.5cm \epsffile{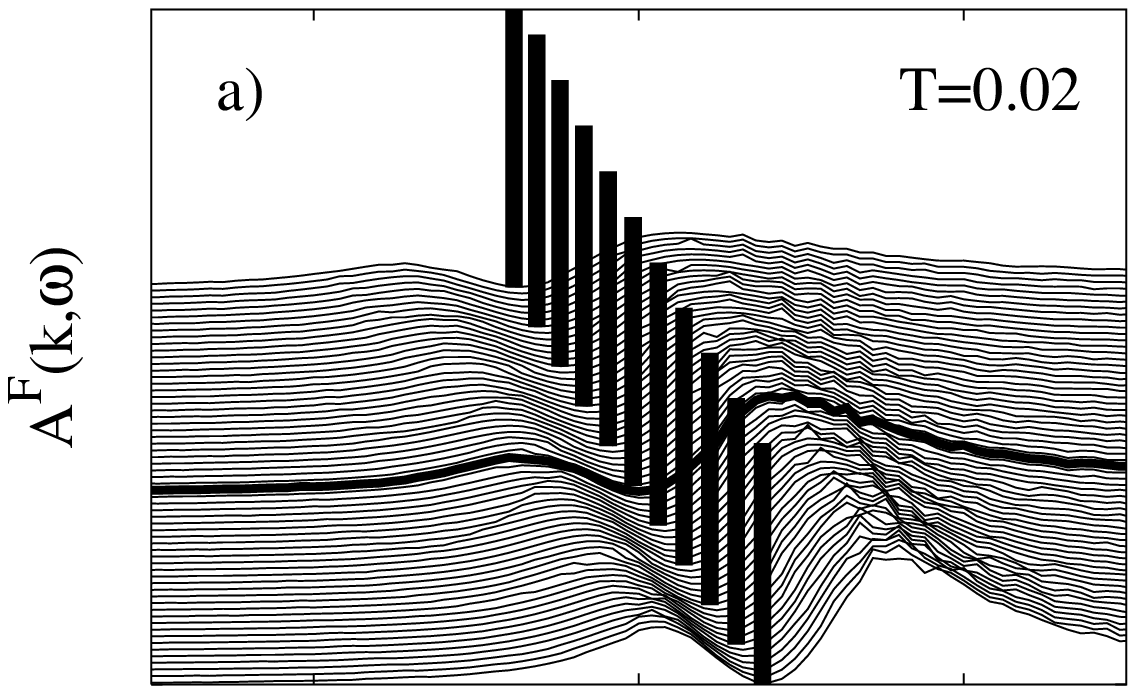}}
\vspace{-13mm}
\centerline{\epsfxsize=8.5cm \epsffile{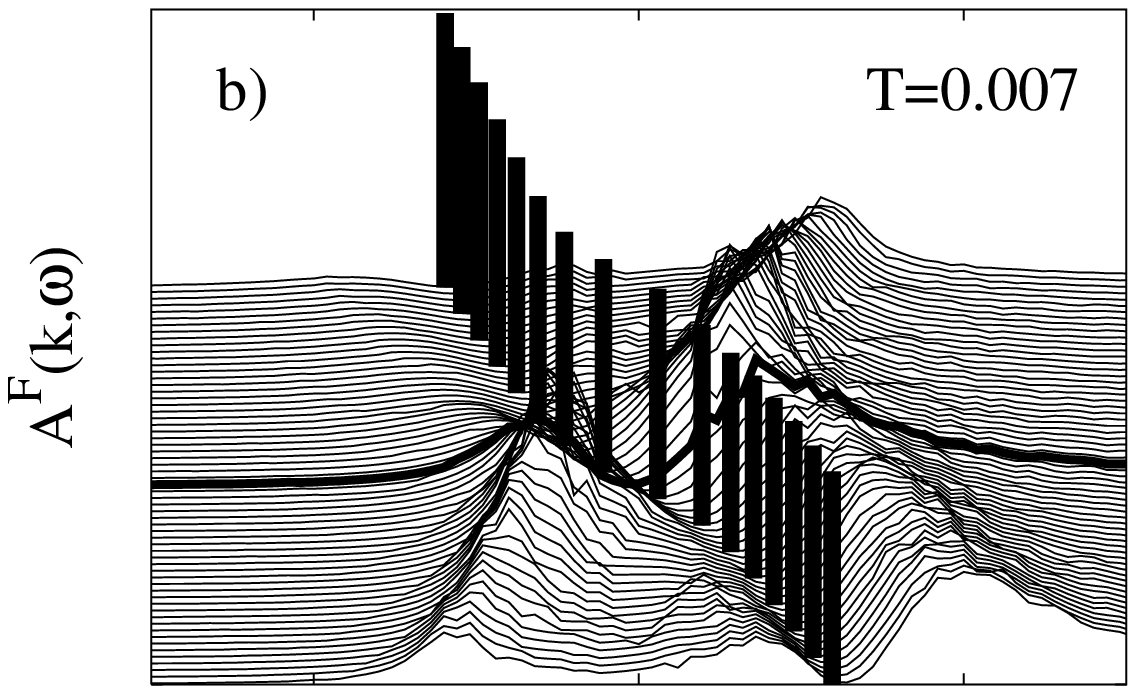}} 
\vspace{-13mm} 
\centerline{\epsfxsize=8.5cm \epsffile{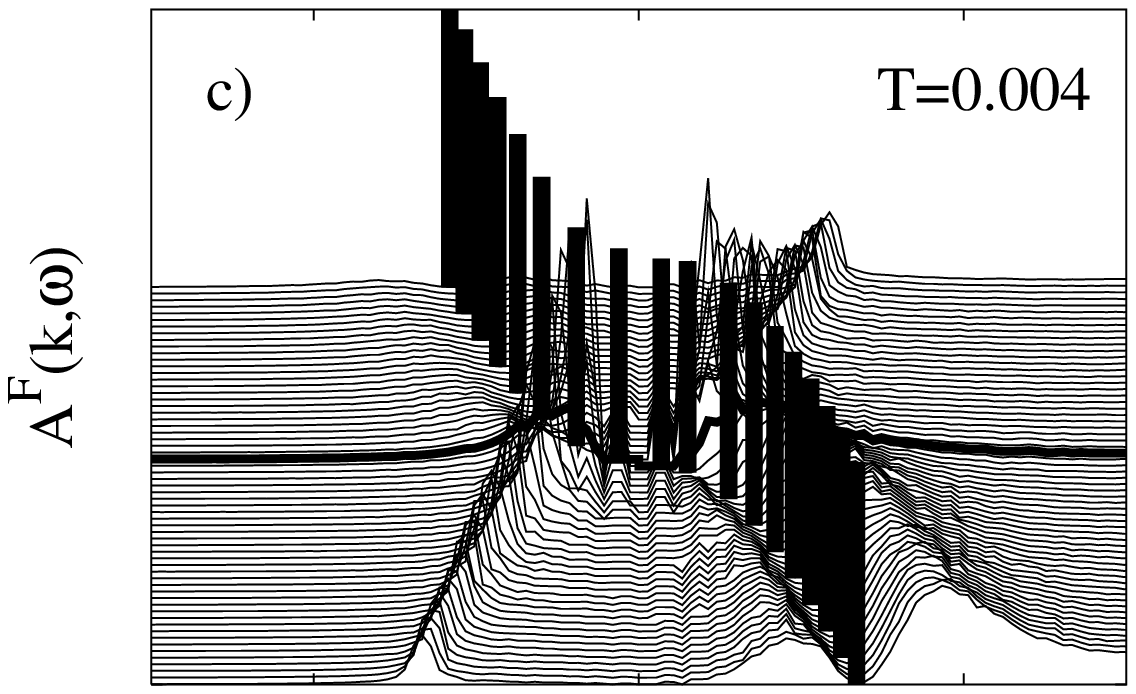}} 
\vspace{-13mm} 
\centerline{\epsfxsize=8.5cm \epsffile{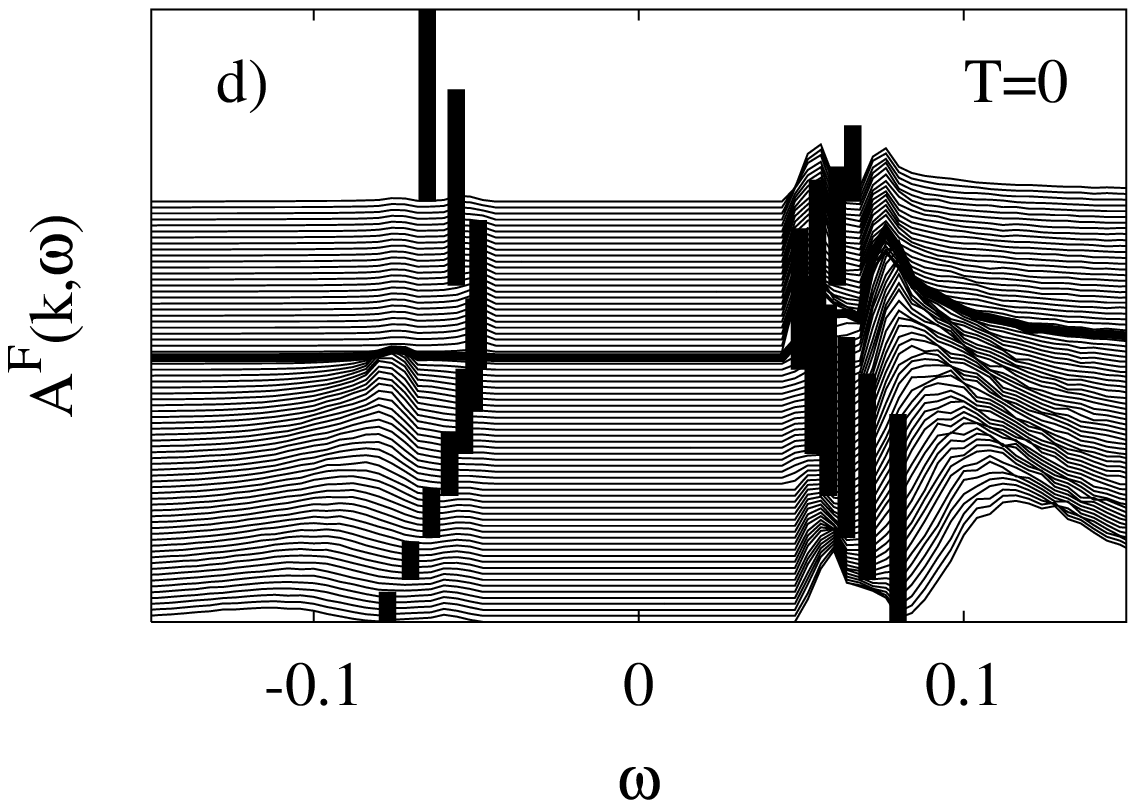}}
\caption{ 
The single particle fermion spectral function
$A^{F}({\bf k},\omega)$ decomposed into its coherent (thick bars whose 
height indicate the intensity of the delta like contributions) sitting on 
top of an incoherent component in the vicinity of ${\bf k_{F}}$
(indicated by the bold spectral line) for the normal phase (a) above $T^{*}$ 
($T=0.02$), (b) and (c)for the pseudogap region $T^{*}>T>T_{c}$ 
($0.007$, $0.004$) and (d) for the superconducting phase (in the ground 
state $T=0$). The distance between the neighboring lines corresponds to 
changes in wavevector by multiples of $\Delta k = \pi /1000a$.}
\label{Fig.1}
\end{figure}

In the normal phase we have $n_{cond}^B = 0$, which implies 
${\cal R}_{\bf k}(l) = p_{\bf k,q}(l) = 0$ for any $l$ and hence
\begin{eqnarray} 
A_{coh}^{F}({\bf k},\omega) &=& 
| {\cal P}_{\bf k}(\infty) |^{2} \delta \left(
\omega - \tilde{\varepsilon}_{\bf k} \right)
\label{Acoh_aboveTc} \\ 
A_{inc}^{F}({\bf k},\omega) & = & 
\frac{1}{N} \sum_{{\bf q}\neq{\bf 0}} \left(
n_{\bf q}^{B} + n_{{\bf q}-{\bf k}\downarrow}^{F}
\right) | r_{{\bf k},{\bf q}}(\infty) |^{2} 
\nonumber \\ & & \times \;
\delta ( \omega - \tilde{E}_{\bf q}
+ \tilde{\varepsilon}_{{\bf q}-{\bf k}} )
\label{Ainc_aboveTc} 
\end{eqnarray}
for the coherent and incoherent contributions to the Fermion spectral 
function. As discussed in ref. \cite{Domanski-01}, for temperatures above a 
certain $T^*$ the Fermion dispersion is essentially unrenormalized and given 
by exclusively $A_{coh}^{F}({\bf k},\omega)$ with 
${\cal P}_{\bf k} \simeq 1$.   
Below $T^*$, on the contrary, the Fermion renormalization is becoming 
increasingly important and gives rise to the opening of a pseudogap in 
the single particle density of states with a spectral weight ${\cal}$
becoming smaller than unity and a consequent redistribution of this lacking 
spectral weight into an incoherent part given by $A_{inc}^{F}({\bf k},\omega)$.
The fully selfconsistent solutions to this problem by numerical means are 
presented in Fig.\ref{Fig.1}. Throughout this work we take as energy unit 
the Fermion bandwidth $2 z t$, ($z$ denoting the coordination number) and the 
initial parameters $\Delta_B= -0.6$ and $v=0.1$. We, moreover, choose
the total particle concentration per site $n_{tot} = n^F + 2 n^B = 1.0$. 
Since we are interested here only in the qualitative aspects of the 
underlying physics and since the physics in the normal state is controlled 
by very local correlations and hence little dependent on the dimensionality, 
we can approximate the various integrals in this temperature regime by their 
sum over a 1D Brillouin zone 
with $2000$ wavevectors, $k_n = n \frac{\pi}{a} \frac{1}{1000}$ and 
$-1000 \leq n \leq 1000$ as long as we don't get too close to $T_c$. 

In the superconducting phase, from inspection of eq. (\ref{f_spectral}), 
we notice the appearance of two contributions to the coherent part of the 
spectral function with spectral weights ${\cal P}_{\bf k}(\infty)^2$ and 
${\cal R}_{\bf k}(\infty)^2$, with a total weight 
${\cal P}_{\bf k}(\infty)^2 +{\cal R}_{\bf k}(\infty)^2$ amounting  to 
less than unity because of the sum rule, Eq. (\ref{PQ_inv}). Well below $T_c$ 
the quasi particle energies are given by\cite{Domanski-01} 
$\tilde{\varepsilon}_{{\bf k}} = \mbox{sgn}(\varepsilon_{\bf k}-\mu) 
\sqrt{(\varepsilon_{\bf k}-\mu)^{2}+v^{2}n_{cond}^{B}}$ and present the 
standard Bogoliubov modes which follow from a straight forward mean field 
analysis of this BFM.
The characteristic features of the Fermion spectral function in the 
superconducting phase can be assessed form their zero temperature 
limit. There, $n_{\bf q}^B =0$ for any ${\bf q} \neq 0$ and the Fermion 
distribution function reduces to a step function 
$n_{{\bf k}\sigma}^F=\theta (-\tilde{\varepsilon}_{\bf k})$. Hence, 
\begin{eqnarray}  
A_{inc}^{F}({\bf k},\omega)&=&\frac{1}{N} 
\sum_{{\bf q} \neq {\bf 0}} \left[ 
| p_{{\bf k},{\bf q}} (\infty) |^{2}
\delta ( \omega + \tilde{E}_{\bf q}
-  \tilde{\varepsilon}_{{\bf q}-{\bf k}})  
\theta (-\tilde{\varepsilon}_{{\bf k}-{\bf q}})\right.  \nonumber \\  
& &\left. +| r_{{\bf k},{\bf q}} (\infty) |^{2}
\delta ( \omega - \tilde{E}_{\bf q}
+  \tilde{\varepsilon}_{{\bf q}+{\bf k}}  )
\theta (-\tilde{\varepsilon}_{{\bf q}+{\bf k}} )
\right]
\end{eqnarray}
In the spirit of the qualitative study, presented here, the sum over the wave 
vectors ${\bf q} \neq 0$ in this expression for $A_{inc}^{F}({\bf k},\omega)$ 
can again be safely approximated by a sum over a $1 D$ Brillouin zone. 

In Figs. 1a-1d we illustrate the evolution with temperature $T$ of the 
coherent (the $\delta$-function like peaks) and the incoherent components 
(the broad hump like features) of the Fermion spectral function in the 
relevant wave vector regime around ${\bf k_F}$ where qualitative changes 
are manifest. In the high temperature 
regime ($T > T^*$, Fig. 1a) the single Fermion spectrum is practically 
unrenormalized, but is constraint to the minimum of a broad ``high 
temperature incoherent contribution'' which is visible 
for energies above as well as below this 
coherent contribution. As we go, upon lowering $T$, into the pseudogap phase 
($T_c < T < T^*$, Figs. 1b, and 1c), an additional  ``low temperature 
incoherent component'' emerges, having a dispersion opposite to that of 
the ``high temperature incoherent contribution''. That latter remains 
practically unchanged all the way down in temperature, and right into the 
superconducting phase. The  ``low temperature incoherent component'', on the 
contrary, noticeably narrows upon decreasing $T$, and eventually strongly 
modifies the behavior of the coherent component,  
foreshadowing  Bogoliubov modes of the superconducting phase.  The spectral 
shape of this low temperature incoherent component'' becomes increasingly 
better defined, as we approach the superconducting phase. Below the pseudogap 
it shows a dispersion which bends downwards for increasing wavectors above 
${\bf k_F}$. Above the pseudogap this dispersion of the shadow Bogoliubov 
mode bends upwards for decreasing wavectors below ${\bf k_F}$. 
These features are clearly  apparent from a comparison of Fig. 1c with Fig. 1d 
for the superconducting phase at $T=0$. In Fig. 1d the 
standard well defined Bogoliubov modes are clearly visible.

In order to highlight this emergence of the Bogoliubov modes out of 
the incoherent contribution of the Fermion spectral function in the normal 
state we illustrate in Fig. 2 the Fermion spectral function at ${\bf k_F}$ 
for different temperatures. We notice the gradual narrowing of 
this low temperature incoherent contribution to this spectral function 
as we approach the superconducting phase upon lowering the 
temperature, with peak positions practically being independent on temperature.

In conclusion, we have shown that in systems with precursor pairing, such as 
given by the Boson Fermion scenario, we can expect rather well defined 
remnants of the Bogoliubov modes in the normal phase in a restricted regime 
above $T_c$. These modes eventually broaden into intrinsically incoherent 
contributions of the spectral function as the temperature increases and 
approaches $T^*$ where resonance pairing of the Fermions ceases. Experimental 
verification of such shadow Bogoliubov bands in the high HTS 
would be decisive in determining whether the pseudogap and 
the superconducting gap in these materials are of common nature or not. 
In the present calculation we neglected the effect coming from the two-body 
interaction $U_{{\bf k},{\bf p},{\bf q}}(\infty)$ whose magnitude is  
small and of the order $\simeq v^{2}$ \cite{Domanski-01}. Nevertheless 
in the pseudogap frequency regime this interaction would lead to a 
shifting away of the quasiparticle peaks from this frequency region (and 
hence reenforce the pseudogap features qualitatively) and simultaneously 
lead to a broadening of the delta peak structure of the coherent 
contributions of order of the Cooperons bandwidth 
$\simeq v^{2}$. This  will be discussed in detail in some future study.

\begin{figure}
\centerline{\epsfxsize=7cm \epsffile{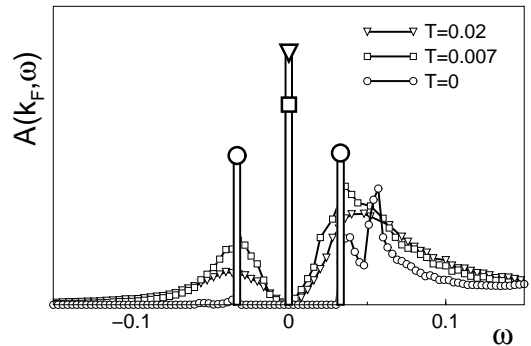}}
\caption{ 
Evolution with temperature of the Fermion spectral function 
$A^{F}({\bf k_F},\omega)$ at ${\bf k_F}$ in the vicinity of the Fermi energy. 
The spectral weights of the $\delta$-function like peaks (corresponding to 
the coherent components sitting on top of the incoherent ones) are indicated 
by squares, circles and triangles.}
\label{Fig.2}
\end{figure}


\begin{thebibliography}{11}
\bibitem{Tchernyshyov-97}
    O.\ Tchernyshyov, Phys.\ Rev.\ B {\bf 56}, 3372 (1997).
\bibitem{Ranninger-02}
    J.\ Ranninger and A.\ Romano, Phys.\ Rev.\ B {\bf 66}, 
    94508 (2002).
\bibitem{Auerbach-02}
    E.\ Altman and A.\ Auerbach, Phys.\ Rev.\ B {\bf 65}, 
    104508 (2002).   
\bibitem{Schnell-99}
    A.\ Schnell, G.\ Roepke and P.\ Schuck, Phys.\ Rev.\ Lett.\
    {\bf 83}, 1929 (1999).
\bibitem{Domanski-03}
    T.\ Doma\'nski, Phys.\ Rev.\ A {\bf 68}, 013603 (2003).      
\bibitem{Timmermans-99}
    E.\ Timmermans, P.\ Tommasini, M.\ Hussein and A. Kerman,
    Phys.\ Rep.\ {315}, 199 (1999).
\bibitem{Chiofalo-02}
    M.L.\ Chiofalo, S.J.J.M.F.\ Kokkelmans, J.N.\ Milstein and 
    M.J.\ Holland, Phys.\ Rev.\ Lett.\ {\bf 88}, 090402 (2001).
\bibitem{Ohashi-03}
    Y.~Ohashi and A.~Griffin, Phys.\ Rev.\ Lett.\ {\bf 89},
    130402 (2002). 
\bibitem{Yurovsky-03}
    V.A.\ Yurovsky and A.\ Ben-Reuven, Phys.\ Rev.\ B
    {\bf 67}, 43611 (2003).
\bibitem{Robin-98}
    J.M.\ Robin, A.\ Romano and J.\ Ranninger, 
    Phys.\ Rev.\ Lett.\ {\bf 81}, 2755 (1998).
\bibitem{Devillard-00}
      P.\ Devillard and J.\ Ranninger, Phys.\ Rev.\ Lett.\ 
    {\bf 84}, 5200 (2000).      
\bibitem{Wegner-94}
    F.\ Wegner, Ann.\ Physik {\bf 3}, 77 (1994).
\bibitem{Glazek-94}
    S.D.\ G\l azek and K.G.\ Wilson, Phys.\ Rev.\ D
    {\bf 84}, 5863 (1994).
\bibitem{Domanski-01}
    T.\ Doma\'nski and J.\ Ranninger, Phys.\ Rev.\ B {\bf 63}, 
    134505 (2001).  
\end{thebibliography}

\end{document}